\documentstyle[aps,multicol,prb,epsf]{revtex} 
 
\begin{document} 
\draft

\title{Effect of dephasing on mesoscopic conductance fluctuations in quantum 
dots with single channel leads} 
 
\author{Edward McCann$\sp{1}$ and Igor V.\ Lerner$\sp{2,3}$} 
\address{$\sp{1}$ Max-Planck-Institut f\"{u}r Physik komplexer Systeme,
{N\"othnitzer Str. 38}, 01187~Dresden, Germany}
\address{$\sp{2}$ School of Physics and Space Research, University of  
Birmingham, 
Birmingham~B15~2TT, United Kingdom 
} 
\address{$\sp{3}$ {Isaac} Newton Institute for Mathematical Sciences, 
Cambridge~CB3~0EH, United Kingdom 
} \date{\today} 
\maketitle \begin{abstract} {
We consider the distribution of conductance fluctuations in 
disordered quantum dots with single channel leads.
 Using a perturbative diagrammatic approach, valid for continuous level
spectra, we describe dephasing due to processes within the dot by considering
two different contributions to the level broadening, thus satisfying particle 
number conservation.
Instead of a completely non-Gaussian distribution, which occurs for zero
dephasing, we find for strong dephasing
that the distribution is mainly Gaussian 
with non-universal variance and non-Gaussian tails.
}\end{abstract} 

\ifpreprintsty\tightenlines \else \begin{multicols}{2} \fi

\bibliographystyle{simpl1}

\section{Introduction}
Soon after the theoretical prediction of universal conductance fluctuations
(UCF) in disordered electronic samples,$^{1-4}$ it
was shown that their distribution function is mainly Gaussian.\cite{AKL:86}
These considerations referred to a weakly disordered {\it open} sample
connected to a reservoir by broad external contacts. 
In this case, after diffusing through the sample of size $L$
during time $\tau_{\text{erg}}\sim L^2/D$ electrons are inelastically
scattered in the reservoir ($D$ is the diffusion coefficient). The resulting
uncertainty in the level position, i.e.\ level broadening $\gamma$, is of the
order of the Thouless energy $E_c=\hbar/\tau_{\text{erg}}$, and
$E_c\gg\Delta$, where $\Delta$ is the mean level spacing. This picture remains
almost unchanged in the presence of relatively strong inelastic processes
within the sample\cite{AlKh:85,Lee:87} which lead to dephasing (the loss of
phase-coherence) at the time scale $\tau_\phi\alt\tau_{\text{erg}}$: the
conductance distribution remains almost Gaussian although its variance is no
longer universal but decreases $\propto (\tau_\phi/\tau_{\text{erg}})^{2-d/2}$
where $d$ is a spatial dimensionality. 

In contrast, it has been shown more recently that the conductance distribution
is pronouncedly non-Gaussian in almost closed systems like chaotic
cavities with single channel leads \cite{Bar:94,Efe:95}, quantum dots in the
regime of Coulomb blockade\cite{Jal:92,Pri:93} where the distribution of the
heights of the Coulomb blockade peaks has been measured
experimentally,\cite{Cha:96} or isolated conducting rings threaded by an
Aharonov--Bohm flux.\cite{KamGef:95} Weak transmission through the contacts
means that the electrons typically spend more time in the system than
$\tau_{\text{erg}}$ so the broadening of energy levels $\gamma \ll E_c$. This
inequality corresponds to the ergodic regime which allows the use of
non-perturbative techniques including random matrix theory
\cite{Meh:91,Bee:97} and the zero dimensional supersymmetric $\sigma$
model.\cite{Efe:83} 

There are two distinct types of behavior within the ergodic regime, depending
on whether the level broadening $\gamma$ is smaller or larger than the mean
level spacing $\Delta$. In the absence of level overlapping, $\gamma \ll
\Delta$, the conductance distribution is clearly non-Gaussian. The nature of
the distribution changes considerably when dephasing is substantial,
$\gamma\agt\Delta$, so that the energy levels overlap. This case allows also a
perturbative treatment\cite{M+L:96} in the framework of the standard
diagrammatic expansion. A crucial question here is how to describe the level
broadening in quantum dots with point-like external contacts. The first
approach, following Ref.~\onlinecite{But:86}, was to attach an additional
voltage probe in which electrons lose their phase coherence before being
re-injected into the dot.\cite{B+M:95} Another approach was to include an
imaginary potential in the Hamiltonian thus allowing for the possibility that
electrons may leave the dot.\cite{Efe:95,Pri:93,M+L:96} However, Brouwer and
Beenakker \cite{B+B:97} have recently pointed out that these approaches do not
describe inelastic processes {\it within} the dot. While the voltage probe
accounts for spatially localized dephasing only, the imaginary potential in
the Hamiltonian does not conserve the particle number. In order to describe
dephasing processes within the dot, they have introduced\cite{B+B:97} in the
framework of random matrix theory a new variant of the voltage probe model
which produces spatially uniform dephasing. Thus it was found that the
conductance distribution became Gaussian for strong dephasing, $\tau_\phi \ll
\hbar/\Delta$. Note that experiments measuring conductance fluctuations in
quantum dots, for example as a function of shape,\cite{Cha:95} also find a
Gaussian distribution.

In this paper we present a microscopic diagrammatic calculation of the
conductance distribution function within a dirty quantum dot with two point
contacts in the presence of strong dephasing processes ($\tau_\phi \ll
\hbar/\Delta$) within the dot. We construct a particle-conserving model by
clearly separating two different mechanisms {of level} broadening: 
dephasing processes which contribute only to the conductance fluctuations and
{particle} leakage which contributes also to the mean value of conductance.
It enables us to resolve a contradiction between the results of the imaginary
potential model \cite{Efe:95,Pri:93,M+L:96} that suffered from the violation
of particle conservation and the voltage-probe model.\cite{B+M:95,B+B:97} By
calculating the moments of the conductance distribution, we will show that 
for $\tau_\phi \ll \hbar/\Delta$ it is almost Gaussian with the variance
\begin{equation}
\text{var}\,G
= \frac{4\zeta_{\phi}(\ell )}{\beta}\,\left<G\right>^2
.  \label{var}
\end{equation}
This expression is valid both in the ergodic zero-mode regime, $\tau_\phi \gg
\tau_{\text{erg}}$, and in the diffusive regime,  $\tau_\phi \ll
\tau_{\text{erg}}$.
Here $\left<G\right>$ is the mean conductance, and 
$\zeta_{\phi}(\ell ) \equiv \zeta_{\phi} (R\!=\!\ell)$
where the small parameter  $\zeta_{\phi} (R)$ is effectively 
a diffusion propagator at distance $R$ in the presence of strong dephasing,
and $\ell$ is the elastic mean free path.
We will show that this parameter is given by
\begin{eqnarray}
\zeta_{\phi}(R)
\approx  \cases{
  g_0^{-1}\ln(L_{\phi}/R),&  $\tau_\phi\ll\tau_{\text{erg}}, \quad d=2$\cr
(\pi /2g_0)\,(\ell/R),& $\tau_\phi\ll\tau_{\text{erg}}, \quad d=3 $\cr
\Delta\tau_\phi/\pi ,& $\tau_\phi\gg\tau_{\text{erg}},\quad $ any $d$
}  \label{zr}
\end{eqnarray}
Here $L_{\phi} = (D\tau_\phi )^{1/2}$ is the dephasing length.
Note that although the voltage-probe model\cite{B+B:97} also
predicts a Gaussian conductance distribution, 
the width of the distribution given by Eq.~(\ref{var}) is
different from that obtained in Ref.~\onlinecite{B+B:97}. In contrast to a
cavity with broad leads,\cite{SerFen} (which {corresponds} to a 
multi-channel
case so that the fluctuations {remain} universal)
the variance is non-universal even in the diffusive regime, 
 $\tau_\phi \ll \tau_{\text{erg}}$.

The factor $\zeta_{\phi}(\ell )$ which governs the distribution width
{resembles} the standard
weak-localization parameter. This could be seen from the first line
of Eq.~(\ref{zr}) where $g_0\sim E_c/\Delta $ is a classical
dimensionless conductance of a sample with broad leads. Nevertheless, the origin
of this parameter is quite different: 
the variance  is not very sensitive to the presence of a magnetic
field and thus has nothing to do with weak localization effects.
When time-reversal {invariance} is absent, the variance (\ref{var})
 changes only by the standard
 Dyson's  factor of $1/\beta$, typically for mesoscopic effects,
where $\beta=1$ for the
orthogonal ensemble (in  the presence of potential
 scattering only), and $\beta=2$ for the unitary ensemble
(in the presence of a finite magnetic field or weak scattering by
magnetic impurities that breaks time-reversal symmetry).

It is worth noting that the close-to-Gaussian 
nature of the distribution is
established in a way somewhat less straightforward than for UCF in
samples with broad leads.\cite{AKL:86} Had it been possible to restrict
considerations to the lowest order of perturbation for the variance and
higher moments, we would have obtained
 the exponential distribution as in the
imaginary potential model, \cite{Efe:95,Pri:93,M+L:96} albeit with a
shifted position of the mean. However, the leading contribution to the
moments is given by the next (two-loop) order of perturbation while
contributions from higher orders are negligible.

\section{Two kinds of diffusons}

In considerations of conductance fluctuations {in quantum} dots with
single-channel leads,
the lowest order of perturbation does not make a leading
contribution due
to  the existence
of two different perturbative parameters related {to qualitatively}
different contributions to the level broadening. Although both of them
 arise also in considering any mesoscopic effect for a diffusive
sample with broad contacts, only one of them is relevant in that case
as we explain below. 

 We adopt the impurity diagram technique  \cite{Abr:65}
to perform ensemble averaging over disorder with allowance
for dephasing processes. At the first step, we assume a conventional
model of free electrons in a random Gaussian 
potential $V$, with $\left<V(r)\right>=0$ and
\begin{equation}
\left<V(\bbox{r})\,V(\bbox{r^\prime})\right>=
\frac1{2\pi\nu_0\tau_{\text{el}}}\,\delta(\bbox{r}-\bbox{r^\prime})\,,
 \label{RP}
\end{equation}
where $\nu_0$ is the one-electron density of states and
$\tau_{\text{el}}$ is the mean elastic scattering time.
 A standard perturbative expansion \cite{Gor:79} is
a loop-expansion with the loops being 
 built up of diffusons and Cooperons (if time reversal symmetry
is preserved). These are impurity ladders (Fig.~1) with a pole 
at small transferred momenta (diffusons) or small total momenta
(Cooperons). 
 For an isolated $d$-cubic  sample of size $L$ with 
$L\gg\ell$
such a diffusion propagator
({\em diffuson}) at $T=0$, i.e.\ in the absence of inelastic
processes, is given by 
\begin{equation}
{\cal D}({\bf q};\omega) = \frac{1}{2\pi\nu_0\tau_{\text{el}}^2}
\frac{1}{D\bbox{q}^2 - i\omega}\,,
 \label{DP}
\end{equation}
where $D=v_F^2\tau_{\text{el}} /d$ is the diffusion constant, $v_F\tau
_{\text{el}}=\ell$, 
 and
$\bbox{q} = \pi \bbox{n}/L$ with
$\bbox{n}=(n_1,\ldots, n_d)$ being non-negative integers.
In the presence of time reversal symmetry,
the Cooperon propagator has the same form.

{For weak} disorder,
the diffusion propagator is proportional to the Fourier transform
of the density-density correlation function, 
\begin{equation}
K({\bf r}; t)\equiv\left<\,\rho(0,0)\,\rho(\bbox{r}, t)\,\right>\,,
 \label{CF}
\end{equation}
which describes the probability of finding 
at the point $\bbox{r}$ at time $t$ a
diffusive particle injected at the origin at $t=0$.
Particle conservation implies that $\int K({\bf r}; t) d^dr=
\text{const}$ at any time $t\ge0$. On performing the Fourier transform,
one then obtains that 
${\cal D}(0;\omega) =\text{const}/\omega$. 

This exact identity remains valid also at $T\ne0$ when one includes a weak
interaction leading to dephasing  processes
in the sample. Therefore, the diffuson associated with
$K({\bf r}; t)$ remains the same as in Eq.\ (\ref{DP}):
the only possible effect of dephasing could be in
replacing  $D$ with $ D_{\text{eff}}$ (for weak disorder,
$|D_{\text{eff}} -D|\ll D$ so that this effect may also be neglected).

\ifpreprintsty \else
\begin{figure}
\vspace{0.3cm}
\hspace{0.02\hsize}
\epsfxsize=0.9\hsize
\epsffile{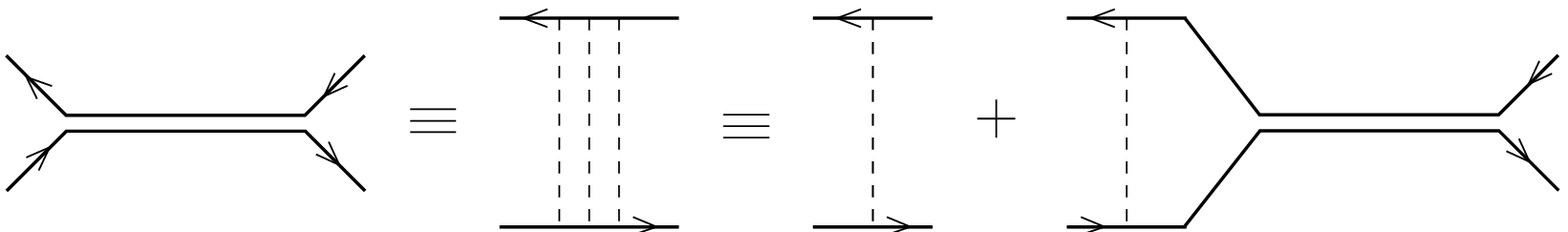}
\vspace{0.3cm}
 
\refstepcounter{figure}
\label{fig:1}
{\small \setlength{\baselineskip}{10pt} FIG.\ \ref{fig:1}.
A diffuson ladder. A Cooperon ladder is obtained by inverting 
the direction of one of {the} arrows. }
\end{figure}
\fi

Mathematically, particle conservation in the diffuson associated
with the density-density correlation function (which we call
{\it the intrinsic diffuson}) is due to the Ward identity
which provides an  exact cancellation of the one-particle
self-energy corrections due to inelastic processes
by vertex corrections (described graphically as interaction lines between
the two sides of the diffusion ladders).
However, if one defines the diffuson in terms of the sum of graphs describing
diffusive propagation in the particle-hole channel {\it without
energy exchange} between the particle and hole line,
it is no longer proportional to
 the density-density correlation function and it is not constrained by
particle conservation. \cite{CdCKL:86}
The absence of the usual cancellation of self-energy corrections
by vertex corrections leads to
the appearance of the cutoff $\tau_\phi^{-1}$ in the diffusion pole.
\cite{Lee:87,CdCKL:86}
 This is precisely the kind of diffuson arising in typical lowest-order
conductance-fluctuations diagrams$^{1-5}$
and describing correlations between {\it different
members of the ensemble} which makes energy exchange between the opposite
sides of the diffusion ladder impossible. We will call this kind 
of diffuson {\it inter-sample}.  Note that all the Cooperon propagators,
either intrinsic
or inter-sample, have exactly the same cutoff, $\tau_\phi^{-1}$,
as the inter-sample diffusons.\cite{Lee:87}

 Therefore, one can allow for
dephasing by merely including  $\tau_\phi^{-1}$ 
as an infrared cutoff in the denominators of {\it inter-sample}
diffusons, i.e.\ substituting into Eq.\ (\ref{DP})  
$\tau_\phi^{-1}$ for $-i\omega$ at zero frequency.
At $T\ne0$, averaging over energies leads\cite{AlKh:85,Lee:87} to 
the saturation of the denominators of {\it inter-sample}
diffusons.
A consistent description of both thermal smearing and dephasing 
requires the use of the temperature technique\cite{Abr:65} and  
the development of a 
microscopic theory of dephasing in the partially open dot 
which would {provide a} dependence of $\tau_\phi$ 
on $T$ and microscopic characteristics of the dot.
Such a theory is well established for relatively large
diffusive samples with broad leads\cite{Sch:74}
or for closed quantum dots.\cite{Siv:94}

However, our aim here is just to describe how dephasing and 
thermal smearing influence mesoscopic 
fluctuations of conductances of the dot. To this end, all we
need is to distinguish between the two kinds of diffusons:
{\it inter-sample}
diffusons which are saturated at $q=\omega=0$, and {\it intrinsic}
diffusons which diverge at $q=\omega=0$. Since the divergence
of this sort could not be treated within the diagrammatic approach,
we introduce  a weak level broadening $\gamma_{\text{esc}}$
due to escape from the dot which provides a cutoff for the
{\it intrinsic} diffusons.  The cutoff in the intrinsic diffuson
 violates the particle conservation law but
since we restrict our considerations to the case 
$\gamma_{\text{esc}}\ll\text{min} (T,\,\hbar/\tau_\phi)$,
it is dephasing (or thermal smearing) within the dot rather than particle
escape which governs the conductance distribution.
Even with this cutoff, ${\cal D}_\phi(q\!=\!\omega\!=\!0) \ll
{\cal D}_{\text{esc}}(q\!=\!\omega\!=\!0)$, i.e.\ there is 
a sharp distinction between the {\it intrinsic}
diffusons, ${\cal D}_{\text{esc}}$, and the {\it inter-sample}
diffusons, ${\cal D}_\phi$.  For simplicity, we shall
use a zero-temperature diagrammatic technique. 
If $T>\hbar/\tau_\phi$, one should substitute $T$ for $\hbar/\tau_\phi$
and the thermal-smearing length $L_T=(\hbar D/T)^{1/2}$ for 
the dephasing length $L_\phi$ in all final results, including that 
for the variance, Eq.~(\ref{var}). 

The existence of the two kinds of diffusons was practically irrelevant 
for the description of mesoscopic conductance fluctuations in samples
with broad leads.$^{1-5}$
Firstly, there is  strong leakage through the leads, 
$\gamma\sim \hbar/\tau_{\text{erg}}$, and 
one could restrict considerations to the case of  weak 
dephasing, $\tau_{\text{erg}}\ll\tau_\phi$, when both intrinsic and
 inter-sample diffusons have the same cutoff.
Moreover, even for larger samples or higher temperatures when
$\tau_{\text{erg}}\gg\tau_\phi$, the intrinsic 
diffusons do not appear either in $\langle G\rangle$ or
in var$G$ but as a third-order perturbation correction which is irrelevant
in a weakly disordered metal. In contrast to this, in the problem under 
consideration the intrinsic diffusons determine the value of
$\langle G\rangle$ and enter var$G$ in the leading
order of perturbation.

 \section{Conditions for perturbative approach}

Traditionally conductance fluctuations  in
 a system with broad, spatially homogeneous 
contacts are considered$^{1-5}$
by means of the Kubo formula \cite{Kub:57}.
Alternative considerations based on 
the Landauer-B\"uttiker formula \cite{Lan:70} are more convenient in
the presence of tunnel barriers \cite{IWZ} or for
a lead geometry which involves spatially inhomogeneous currents. 
In this paper we will 
use this formula in the form first derived by Fisher and Lee:\cite{F+L:81}
\begin{equation}
G = {e^2\over{2h}} \sum_{ab} \left(
T_{ab}^{L}+T_{ab}^{R}
\right)  ,   \label{lb}
\end{equation}
where the transmission coefficient $T_{ab}^{L(R)}$ is the probability of 
transmission from the channel labeled by $a$ in the left (right) lead to 
the channel labeled by $b$ in the right (left) lead.
In the case of single-channel leads (of width $w \simeq \hbar k_F^{-1}$) 
the point-to-point conductance, $G$, is 
given by Eq.\ (\ref{lb})  without the summation sign, 
 the transmission coefficients in this equation 
being related to Green's functions by 
\cite{F+L:81}
 \begin{equation}
T^{L(R)} = \frac{\alpha_1\alpha_2}{(h\nu_0)^2} \Big[{\cal G}^{+(-)}\left( 
{\bf  r_1} , {\bf  r_2} ; \varepsilon \right) 
{\cal G}^{-(+)}( {\bf  r_2},  {\bf  r_1} ; \varepsilon )\Big]\,, 
\label{tab}
 \end{equation}
where ${\cal G}^{+}$ (${\cal G}^-$) is a retarded (advanced) Green's function,
$ {\bf r_1}$, $ {\bf r_2}$ are the positions of the point contacts, and
$\alpha_1\alpha_2$ is the transmission probability through the contacts
themselves. In the entire energy interval of interest the mean density of
states $\nu_0$ is a constant and the $T^{L(R)}$ are energy independent so we
will subsequently drop the $\varepsilon$ label. Thus we imply that energy
dependence will also be irrelevant for the conductance fluctuations: as usual,
this is valid when the energy difference between two conductances, $\omega$,
is much smaller than the level broadening which we assume to be the case.

We consider the region of parameters defined by
\begin{equation}
\Delta\alt\gamma_{\text{esc}}\ll \min\left\{
\hbar/{\tau_\phi}\,,\,T\,,\,
\hbar/{\tau_{\text{erg}}}
\right\}\,,
\label{ineq}
\end{equation}
Here the level broadening due to escape from the dot is represented by
$\gamma_{\text{esc}}\approx\Delta(\alpha_1+\alpha_2 )+
\gamma_{\text{leak}}$ where $\gamma_{\text{leak}}$ is the level broadening
due to escape from the dot other than
through the point contacts, for example due to leakage through the dot matrix.

Our perturbative approach is formally {applicable both} to the non-ergodic
diffusive regime, $\tau_\phi\alt \tau_{\text{erg}}$, and to the ergodic
zero-mode regime, $\tau_\phi\agt \tau_{\text{erg}}$. The inequality $
\gamma_{\text{esc}}\ll \min(\hbar/{\tau_\phi},T)$ is a crucial distinction
from the imaginary potential model \cite{Efe:95,Pri:93,M+L:96} where the
opposite inequality holds. This allows us to separate dephasing processes inside
the dot from those due {to weak particle} leakage. The latter have been
introduced only for convergence of the perturbative approach which is not
valid for $\gamma_{\text{esc}}<\Delta$. Therefore, a weak violation of
particle conservation is inevitable in the perturbative approach as will be
explicitly shown later. However, this weak particle non-conservation is not
related to the level broadening due to dephasing inside the dot and, as we
shall see, may be totally neglected in the final results. On the contrary, in
the imaginary potential model both the level broadening and the particle
non-conservation have exactly the same source and cannot be separated which
makes this model unsuitable for considerations of the influence of inelastic
processes inside the dot on the conductance distribution. 

The nonperturbative region $\gamma_{\text{esc}}<\Delta$ was the main area of
interest for previous zero mode calculations. \cite{Bar:94,Pri:93,Efe:95}
However, the exact zero-mode calculations within the nonlinear $\sigma$ model
cannot be extended (at least, in a straightforward way) to include relaxation
processes inside the dot by simply introducing $\tau_\phi$. The reason is that
all intra-sample diffusons {\it must} and all intrinsic diffusons {\it must
not} contain a cutoff proportional to $\tau_\phi^{-1}$. These two kinds
of diffusons which are both elementary excitations in the nonlinear $\sigma$
model must be distinguished automatically. This can be easily done in the
perturbative approach to the $\sigma$ model \cite{AKL:86} by the introduction
of an extra set of matrix indices to numerate different conductance loops.
While this additional dependence does not lead to any complication in
perturbative or renormalization group calculations, it would make direct zero
mode calculations rather difficult if possible at all. Obviously, one
circumvents this {complication} when dephasing is included 
phenomenologically,
as in the voltage probe model where one can proceed with zero-mode
calculations. \cite{B+M:95,B+B:97} However, there is a price
to pay: as two kinds of diffusons are not {distinguished}, this model 
contains
 $\tau_\phi$
proportional corrections to the mean conductance\cite{B+B:97} even in the
absence of time-reverse invariance when such corrections cannot appear in all
orders of perturbation as we will show below.
 
The present model overlaps with the voltage probe model in the ergodic
zero-mode regime (with strong dephasing): $\tau_{\text{erg}}\ll \tau_\phi\ll
\tau_{\text{esc}}$. Even in this case, the overlapping is restricted to the
region $\gamma_{\text{esc}}\agt \Delta$ as the requirement of convergence of
perturbation series does not allow us to take the limit
$\gamma_{\text{leak}}\to0$.
Nevertheless, both the models address clearly the same physical situation
where dephasing occurs inside the dot, and  the results of the present
considerations are in a broad agreement with the  voltage probe model.
Moreover, we shall argue later that the
perturbative approach might be valid even outside the formal limits of
applicability. Note finally that since our considerations always refer to
{strong dephasing} {with} a noticeable leakage ($\gamma_{\text{leak}}\agt 
\Delta$),
 the results are applicable both for statistics of the
heights of the Coulomb blockade peaks (for $\alpha_{1,2}\ll 1$) and for the
conductance distribution of a single-channel cavity (for $\alpha_{1,2}\alt
1$).

 Conductance cumulants
$\left<\!\left<G^n\right>\!\right>$ are given
 by
\begin{equation}
\left<\!\left<G^n\right>\!\right> = 
\left({e^2}/h\right)^n \left<\!\left<  \left(
T^L+T^R
\right)^n \right>\!\right>,
 \label{lb2}
\end{equation}
where $T^{(L,R)}$ should be expressed in terms of Green's functions,
Eq.~(\ref{tab}), and an extra prefactor
of 2 is due to spin.
Fisher and Lee \cite{F+L:81} have shown that the unitarity of the 
$S$ matrix implies $T^R=T^L$
even in the absence of time reversal invariance
(in \cite{Mar:93} the proof of this equality
for a two lead system has been explicitly based on particle conservation). 
In the imaginary potential model of Refs.~\onlinecite{Efe:95,Pri:93,M+L:96}
the particle number is not conserved and, 
if time reversal symmetry is broken,
$T^R\ne T^L$ as we will show explicitly later.
 That is why the conductance distribution
in the absence of  time reversal symmetry 
obtained\cite{M+L:96} from the
 microscopic formula (\ref{lb}) turns out to be different from that
obtained\cite{Pri:93,Efe:95} from the reduced formula $G=(e^2/h)T^L$.

 \section{Mean conductance}
 
In the model under consideration, the particle number is 
{\it almost} conserved: for the sake of convergence we have
introduced $\gamma_{\text{esc}}\agt\Delta$ which describes particle leakage;
however, in the region of interest, $\gamma_{\text{esc}}\ll\hbar/\tau_\phi$,
the leakage effect is small and we will show that in the leading order
approximation $T^R= T^L$. On average though,
 $\left<T^R\right>= \left<T^L\right>$
so that the calculation of the mean conductance is straightforward.

The mean conductance $\left<G\right>$ is clearly proportional to the
Fourier transform at zero frequency of the
density-density correlation function, Eq.~(\ref{CF}). For the parametric
region of Eq.~(\ref{ineq})
and {for point} contacts separated by a large distance,
$R\equiv|{ {\bf  r_1} - {\bf  r_2}}| \gg \ell$, 
 $\left<G\right>$ is dominated by the one-diffuson contribution 
shown in Fig.~2.

\ifpreprintsty \else
\begin{figure}
\vspace*{0.3cm}
\hspace{0.12\hsize}
\epsfxsize=0.6\hsize
\epsffile{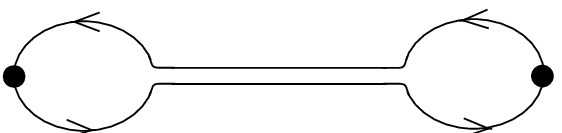}
\vspace{0.3cm}
 
\refstepcounter{figure}
\label{fig:2}
{\small \setlength{\baselineskip}{10pt} FIG.\ \ref{fig:2}.
The diagram for the mean conductance.}
\end{figure}\fi

 Obviously, this is the {\it intrinsic} diffuson,
obeying particle conservation. 
At each end of the diffuson there are two-sided `petal' shapes
given analytically by
\begin{equation}
\chi_2( {\bf  r} , {\bf  r^{\prime}})=
\left<{\cal G}^{+}( {\bf  r} - {\bf  r^{\prime}})\right>
\left<{\cal G}^{-}( {\bf  r^{\prime}} - {\bf  r})\right>
\label{petal}
\end{equation}
The filled circles at the petals  correspond to the constants
 $\alpha_{1,2}/(2\pi\nu_0)$ where $\alpha_{1,2}$ are 
 the transmission probabilities  through the contacts
(here and elsewhere we put $\hbar$ to $1$ in all intermediate 
expressions). 
The petals correspond to motion at ballistic scales since the average Green's
functions $\left<{\cal G}^{\pm}( {\bf  r} , {\bf  r^{\prime}})\right>$
(drawn as edges of the petal) decay like $\exp (-| {\bf  r} -
{\bf  r^{\prime}} | /2\ell )$. 
 At the diffusive scale, $| {\bf  r} -
{\bf  r^{\prime}}| \agt \ell$, the petals may be approximated by
$\chi_2( {\bf  r} , {\bf  r^{\prime}})=
\chi_2\,\delta({\bf  r} - {\bf  r^{\prime}})$ where the constant
$\chi_2 $ is given by\cite{tau-el}
                  \ifpreprintsty
                   \else 
			\end{multicols}\vspace*{-3.5ex}{\tiny
                   \noindent\begin{tabular}[t]{c|}
                   \parbox{0.493\hsize}{~} \\ \hline \end{tabular}}
                                     \fi
\begin{equation}
\chi_2=\int \left<{\cal G}^{+}( {\bf  r} - {\bf  r^{\prime}})\right>
\left<{\cal G}^{-}( {\bf  r^{\prime}} - {\bf  r})\right>
{\rm d}^d r=
\int \left<{\cal G}^{+}( {\bf p})\right>
\left<{\cal G}^{-}( {\bf p})\right>{\rm d}^d p
=\nu_0\int_{-\varepsilon_F}^{\infty}\frac{{\rm d}\xi}{\xi^2+\frac1{4\tau^2}}
\approx 2\pi\nu_0\tau\,.
\label{chi2}
\end{equation}
where $\xi=p^2/2m-\varepsilon_F$. Then one finds
\begin{equation}
\left<G\right> =2A
\int\chi_2( {\bf  r_1} , {\bf  r_1^{\prime}})
{\cal D}({\bf r}_1^{\prime},{\bf r}_2^{\prime};\gamma_{\text{esc}})
\chi_2( {\bf  r_2} , {\bf  r_2^{\prime}})
\,{\rm d}^d r_1^{\prime} {\rm d}^d r_2^{\prime}\, 
={e^2\over{2\pi}}\alpha_{1}\alpha_{2} \zeta_{\text{esc}} (R) \,,
\quad A\equiv
{e^2\over{2\pi}} \frac{ \alpha_{1}\alpha_{2}}{(2\pi\nu_0)^2}\,, 
\label{meang}
\end{equation}
                   \ifpreprintsty
                   \else
		 {\tiny\hspace*{\fill}\begin{tabular}[t]{|c}\hline
                    \parbox{0.49\hsize}{~} \\
                   \end{tabular}}\vspace*{-2.5ex}\begin{multicols}{2}\noindent
                    \fi
where $R\equiv
|{\bf r}_1-{\bf r}_2|\,,$ and  $\zeta_{\text{esc}} (R)$ is defined by
\begin{equation}
\zeta(R)\equiv  {2  \tau^2}
{\cal D}({\bf r}_1,{\bf r}_{2};i\gamma) = \frac{\Delta}{\pi}
\sum_{{\bf q}}  
\frac{e^{ i{\bf q}\cdot({\bf r}_1- {\bf r}_2)}}{D\bbox{q}^2 + \gamma} 
\,,   \label{dr}
\end{equation}
with $\gamma=\gamma_{\text{esc}}$. As $\gamma_{\text{esc}}\ll E_c$
in the parametric regime
(\ref{ineq}),  $\zeta_{\text{esc}} (R)$ is dominated by the
zero-mode contribution ($q=0$) only. Therefore, 
\begin{equation}
\zeta_{\text{esc}}(R) \approx {\Delta\over{\pi\gamma_{\text{esc}}}} \,.
\label{esc}
\end{equation}
Combining Eqs.\ (\ref{meang}) and (\ref{esc}), one finds
\begin{equation}
\langle G\rangle = \frac{e^2}{\pi h}
{\alpha_{1}\alpha_{2}\Delta\over{\gamma_{\text{esc}}}} \,.
\label{G}
\end{equation}
 As expected in the zero-mode approximation, 
  $\left<G\right>$ is independent of the separation of the point 
contacts, the dimensionality, and the degree of disorder.
The perturbation expansion for $\left<G\right>$ is strictly valid only in
the regime where $\zeta_{\text{esc}}\ll1$. Without introducing
$\gamma_{\text{leak}}$, we would have $\gamma_{\text{esc}} \approx \Delta
(\alpha_{1}+\alpha_{2}) \ll\Delta$ so that $\zeta_{\text{esc}}\gg1$.
However, substituting this value of $\zeta_{\text{esc}}$ into Eq.\
(\ref{meang}) reproduces the result of the voltage probe model 
 for the classical mean conductance
\cite{B+B:97} which corresponds to the standard
single-channel conductance.

 This means that all perturbative corrections to
$\left<G\right>$ either vanish or cancel each other. In the absence of time
reversal symmetry, higher order corrections could result from the expansion
in the diffuson loops only. In the usual case of broad contacts, 
 one should sum over all $q$ in Eq.\
(\ref{dr}) which leads to the parameter, $g_0^{-1}\ln L/\ell$
 at $d=2$.
It is well known that all the `main
logarithms', $g^{-n}\ln ^n\omega\tau $,
are mutually cancelled due to the renormalizability in all orders of
perturbation. \cite{Gor:79,Sch+W,Efe:83} The first nonvanishing
contribution arises in the second order and is proportional to
$g^{-2}\ln\omega\tau $ in correspondence with the renormalization group
results. \cite{Sch+W,Efe:83} However, in the zero mode regime considered
here there is complete cancellation at least up to the fourth order
similar to that occurring in the same regime in calculating the
energy-level correlation function\cite{CLS2}. This is easy to verify as the
calculation is much simpler than the standard one in the diffusive regime.
One could assume that the cancellation of the main logarithms in all higher
orders
in the diffusive regime corresponds to a similar cancellation 
 of the zero mode contributions. Since in the latter case there is no
renormalization group backing, such a cancellation remains
only a plausible hypothesis. Nevertheless, it is certain that in
the absence of time-reversal invariance no correction due to dephasing may
appear as they are forbidden by particle conservation. When time-reversal
symmetry is not broken, there exist weak localization (Cooperon)
corrections in the lowest order which are proportional to the standard
$g^{-1}\ln\tau_\phi/\tau$ in the non-ergodic diffusive regime,
$\tau_\phi\ll\tau_{\text{erg}}$, or to $\Delta/\tau_\phi$ in the ergodic
zero-mode regime, $\tau_\phi\gg\tau_{\text{erg}}$. In any case, in the
region of Eq.\ (\ref{ineq}) these corrections are negligible, and the main
result is given by  Eq.\ (\ref{G}).
 
 \section{Variance of conductance}

The lowest order diagrammatic contribution to the variance is given
in Fig.~3, for clarity both in the standard representation 
$^{1-4}$
and in that with diffusion modes separated from
ballistic ones \cite{Gor:79} as in Fig.~1. Although the diagram 
in Fig.~3a looks precisely like one of the leading diagrams$^{1-5}$
for the UCF, the result of the calculation is absolutely
different due to the fact that the vertices of both conductance loops 
correspond to the point contacts.
This is most clear in the representation
of Fig.~3b where local petals are connected together by two inter-sample
{diffusons}.  The petals reduce to the same constant as for the
mean conductance, Eq.\ (\ref{chi2}).
Thus, one finds by analogy with
Eq.\ (\ref{meang}) that this  diagram's contribution to the variance is
$$
2A^2 \chi_2^2 {\cal D}^2
(\bbox{r}_1, \bbox{r}_2;\,i\gamma_\phi)\,,
$$

\ifpreprintsty \else
\begin{figure}
\vspace{0.3cm}
\epsfxsize=0.95\hsize
\epsffile{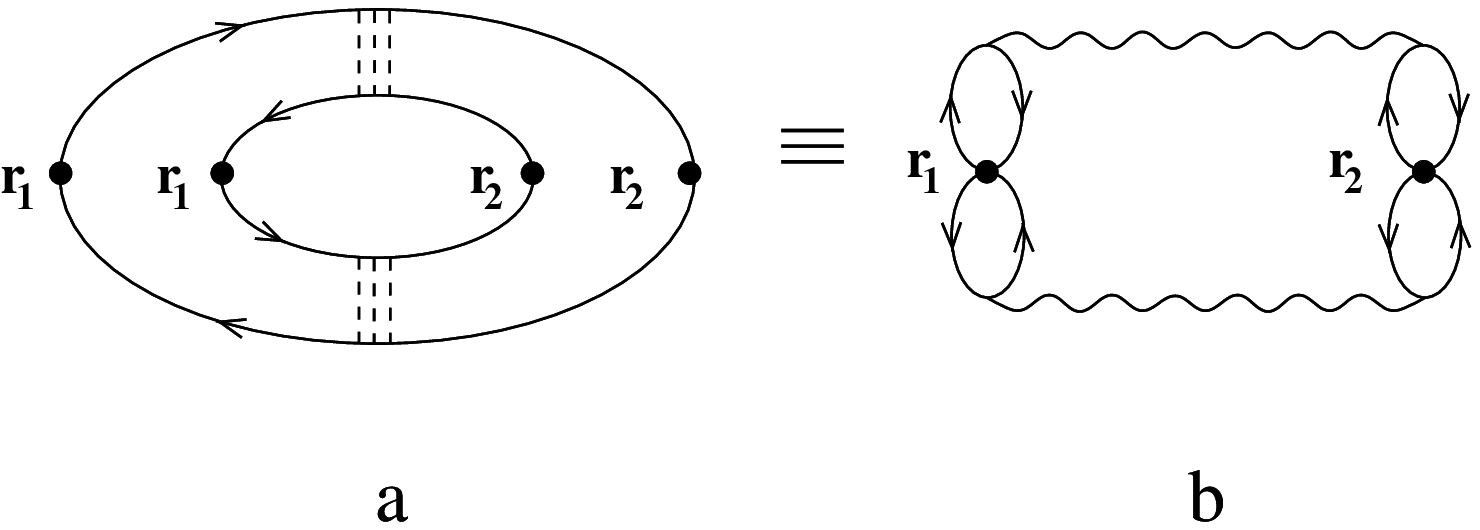}
\vspace{0.3cm}

\refstepcounter{figure}
\label{fig:3}
{\small \setlength{\baselineskip}{10pt} FIG.\ \ref{fig:3}.
The lowest-order contributions to the variance. Wavy lines represent
{diffuson} ladders which correspond to the {inter-sample} diffusons. The
relation between a wavy line and a ladder is the same as for intrinsic
diffusons, Fig.\ \ref{fig:1}.
}
\end{figure}\fi

\noindent 
where the constant $A$ is the overall factor in the expression
for $T^{L,R}$ in terms of Green's function, Eq.~(\ref{tab}), this
factor being explicitly defined  in Eq.\ (\ref{meang}). 
Substituting the values of all constants and taking into account
the contribution of the equivalent Cooperon diagram 
(obtained from that in Fig.~3a
by inverting the direction of arrows in one of the loops)
results in the following expression
\begin{equation}
\left<\!\left<G^2\right>\!\right>^{(1)} =
\frac1\beta\,
\left[{e^2\over{h}} \alpha_{1}\alpha_{2} \, 
\zeta_{\phi}(R)\right]^2
\,
 .  \label{vg1}
\end{equation}
Here $\zeta_{\phi}(R)$
is given by Eq.\ (\ref{dr}) with $\gamma=\tau_\phi^{-1}$.
Its value depends on whether one considers the
non-ergodic diffusive regime, $\tau_\phi\ll\tau_{\text{erg}}$, or the
ergodic zero-mode regime, $\tau_\phi\gg\tau_{\text{erg}}$. In the former
case, the summation in Eq.\ (\ref{dr}) may be approximated by an
integration with cutoff at $q\sim L_{\phi}^{-1}=(D\tau_\phi )^{-1/2}$,
 while in the latter case only 
the $q=0$ term  makes a relevant contribution to the sum.  This leads to
the expression for $\zeta_{\phi}(R)$ given in the Introduction,
Eq.~(\ref{zr}). 
The contribution (\ref{vg1}) to the variance 
 differs from $\left<G\right>^2$ by the substitution of
$\zeta_{\phi}(R)$ for $\zeta_{\text{esc}}(R)$.
As $\zeta_{\phi}(R)\ll\zeta_{\text{esc}}(R)$ in the region
(\ref{ineq}), this difference (which is absent  
 in the  imaginary
 potential model\cite{Efe:95,Pri:93,M+L:96}) is crucial for the 
description of dephasing within the dot.

It is important that the diffuson and Cooperon diagrams 
{originate} from different terms in Eq.~(\ref{lb2}). Indeed, a diffuson ladder
connects ${\cal G}^+$ and  ${\cal G}^-$ lines which have the opposite
directions, while a Cooperon ladder connects  ${\cal G}^+$ and  ${\cal G}^-$
with the same direction.
Therefore, in the diffuson diagram in Fig.~3a
${\cal G}^+$'s have the same direction in both loops. Then it follows
from the expression for $T^{(L,R)}$ in terms of Green's functions,
Eq.~(\ref{tab}), that this 
diagram originates from $\bigl<\bigl(T^L\bigr){}^2\bigr>+
\bigl<\bigl(T^R\bigr){}^2\bigr>$.
 As the direction of 
arrows must be inverted in one of the loops in order to
obtain  an equivalent Cooperon 
diagram, in this diagram ${\cal G}^+$'s have the opposite directions
so that it {originates} from $2\bigl<T^L\,T^R\bigr>$. 
 Considering
\begin{eqnarray}
\bigl<\left(T^L-T^R\right){\!}^2\bigr>=2\bigl<\left(T^L\right){\!}^2\bigr>-
2\bigl<T^L\,T^R\bigr>\,,
\label{L=R}
\end{eqnarray}
where, in this lowest diagrammatic order, the first contribution is given by
the diffuson diagram while the second is given by the 
Cooperon diagram, one can see that $T^L\ne T^R$ for $\beta=2$ 
(in the absence of time reversal symmetry)
when the Cooperon contribution vanishes.
  For $\beta=1$ the two contributions in Eq.~(\ref{L=R})
cancel each other.
In the imaginary potential model\cite{Efe:95,Pri:93,M+L:96}  
 the lowest order diagrams
make the leading contributions to the variance.
Therefore, in this
case the reduced formula $G\propto T^L$ is not equivalent to the
microscopically derived\cite{F+L:81} formula (\ref{lb}).
Since the $T^L=T^R$ equality is based on particle
conservation, its breakdown in the imaginary potential model is another
 clear {indication} of the absence of particle
conservation.

In the present model where the particle number is `almost' conserved, the
 breakdown of this equality in the lowest order indicates that the higher
order diagrams must be more relevant. This is indeed the case {since} 
the condition $\zeta_{\phi}\ll\zeta_{\text{esc}}$ suggests 
that the dominant contribution to the variance in the
region (\ref{ineq}) is given by
diagrams which contain the lowest possible number of
intrinsic diffusons ($\zeta_{\text{esc}}$) and have their conductance loops
connected {\it by only one} inter-sample
diffuson ($\zeta_{\phi}$).

  Two such diagrams are shown in Fig.~4 in the two equivalent
representations described above. Directions of arrows are chosen in such 
a {way} that all propagators in both the diagrams are diffusons. Inverting
the direction of arrows in one of the conductance loops
in diagrams (a) and (c), one converts
inter-sample diffusons into Cooperons, leaving intrinsic diffusons intact.  
In the diffuson diagram (4a),  ${\cal G}^+$'s have opposite directions
in different conductance loops
so that it {originates} from  $2\bigl<T^L\,T^R\bigr>$. On the contrary,
${\cal G}^+$'s have the  same  directions in both loops in the diagram (4c) so
that this diagram originates from $\bigl<\bigl(T^L\bigr){}^2\bigr>+
\bigl<\bigl(T^R\bigr){}^2\bigr>$.
Thus, in contrast to the lowest order diagrams in Fig.~\ref{fig:3}, even
for $\beta=2$ when time-reversal symmetry is broken, both 
 $2\bigl<T^L\,T^R\bigr>$ and  $\bigl<\bigl(T^L\bigr){}^2\bigr>+
\bigl<\bigl(T^R\bigr){}^2\bigr>$ contribute to the conductance.

\ifpreprintsty \else
\begin{figure}
\vspace{0.3cm}
\epsfxsize=0.95\hsize
\epsffile{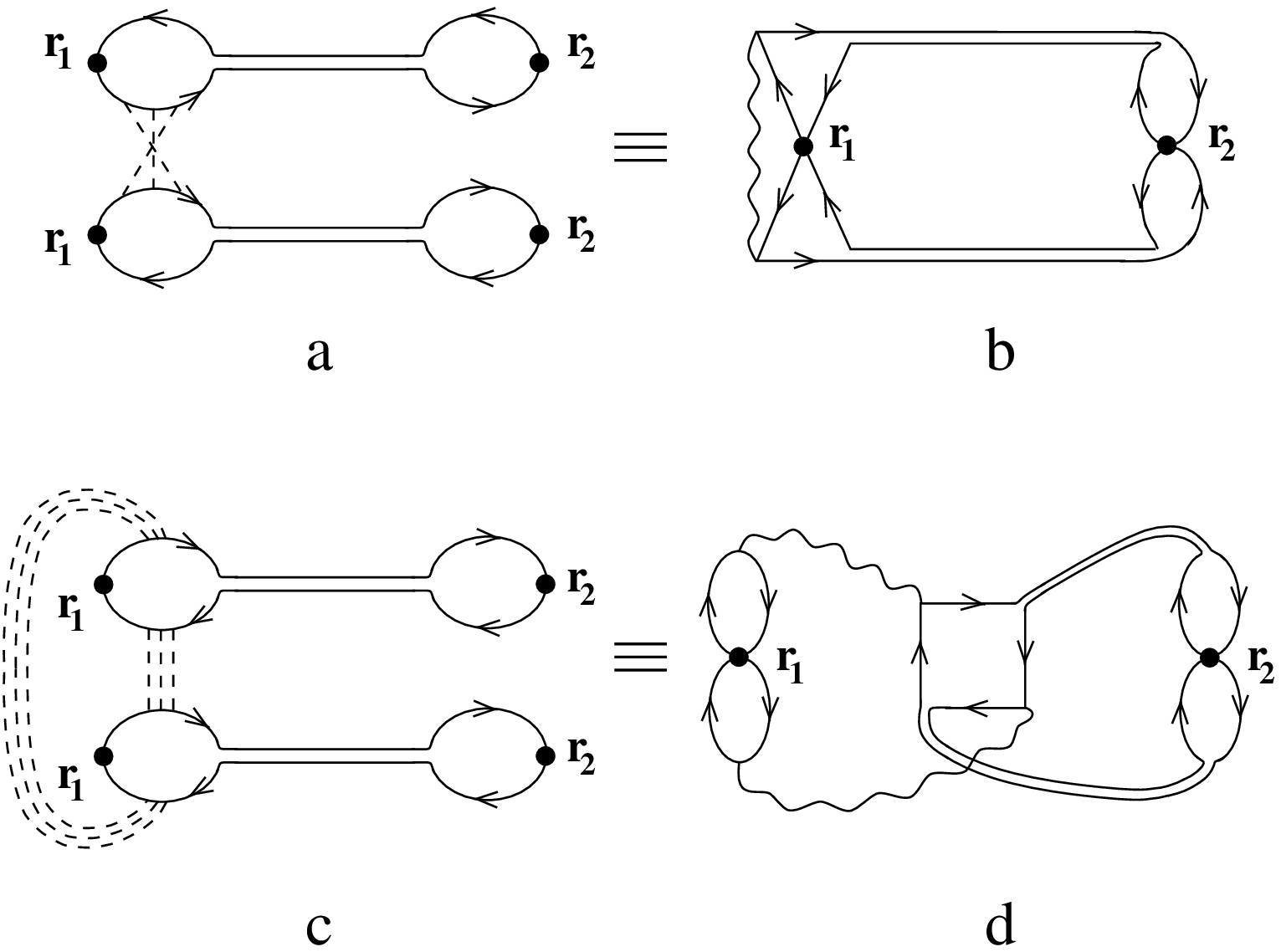}
\vspace{0.3cm}
 
\refstepcounter{figure}
\label{fig:4}
{\small \setlength{\baselineskip}{10pt} FIG.\ \ref{fig:4}.
The leading diagrammatic contribution to the variance. 
In the diagrams {\em b} and {\em d}, the inter-sample diffusons
are represented by wavy lines and the intrinsic diffusons by double
lines.
}
\end{figure}\fi
The calculation of these diagrams is described in the Appendix. 
 For simplicity, we did not draw single-impurity lines
in diagrams in Fig.~4c and 4d which ``dress''
  in a standard way \cite{Gor:79} the inner square in Fig.~4d.
Such a dressing results in an  exact cancellation of one of the
inter-sample diffusons (wavy lines) by the square 
which makes the contribution of the diagram 4d equal to that
of the diagram 4b. In addition to these diagrams, one can 
construct only one more contribution to the variance which 
contains two diffuson loops. This contribution describes a weak-localization
correction to a diffuson and contains three intra-sample ladders thus being
small even compared to the contribution of Eq.~(\ref{vg1}).

Adding together the contributions of all relevant
diffuson and Cooperon diagrams in this order, we find 
\begin{equation}
\left<\!\left<G^2\right>\!\right>^{(2)} =
{4\over{\beta}}\left[{e^2 
\alpha^{2} \over{h}}
\zeta_{\text{esc}}(R)\right]^2 \!\! \zeta_{\phi}(\ell )
\,
.  \label{vg2}
\end{equation}
 Comparing  Eqs.\ (\ref{vg2}) and (\ref{vg1}),
one has 
\begin{equation}\label{ratio}
{\frac{\left<\!\left<G^2\right>\!\right>^{(2)}}
{\left<\!\left<G^2\right>\!\right>^{(1)}}}
\agt\frac{\zeta_{\text{esc}}^2}{\zeta_\phi(\ell)}\gg1
\end{equation}
 in the parametric region (\ref{ineq}) corresponding to strong dephasing
since $\zeta_\phi(\ell)$ is small while $\zeta_{\text{esc}}$ is close to 1.
Therefore, in this regime the variance is
dominated by the non-universal contribution of Eq.~(\ref{vg2}). Substituting
the expression for $\left<G\right>$ into Eq.~(\ref{vg2}), we obtain
the result of Eq.~(\ref{var}) given in the Introduction.

Let us stress that the exact equality of the diffuson diagrams 
(\ref{fig:4}a) originating from $\bigl<T^L\,T^R\bigr>$
and (\ref{fig:4}c)  originating from
$ \bigl<\bigl(T^R\bigr){}^2\bigr>$ 
ensures  that the two terms in Eq.~(\ref{L=R}) cancel each other even
in the case $\beta=2$ when there are no Cooperon contributions. The same
is valid for the leading contribution to the higher order moments
of $T^L-T^R$. Therefore, keeping only the leading contribution
and neglecting sub-dominant
contributions to $\left<\!\left<G^n\right>\!\right>/\left<G\right>^n$
which are due to a small particle 
leakage, one finds that  $T^L=T^R$ as expected in any particle conserving model. 	
 
 \section{{Higher} moments and distribution of  conductance}

In order to determine the distribution, we need to find the leading
contributions to the $n$th cumulant.
The lowest order diagrams are a generalisation of those for 
the variance \cite{M+L:96}, Fig.~3.
 They have two $n$-petal daisy vertices, the petals being
connected by $n$ inter-sample
diffusion propagators.
 Each diagram gives a contribution of $(n-1)!\left<G\right>^n
[\zeta_{\phi}(R)/\zeta_{\text{esc}}(R)]^n$ where
 a factor of $(n-1)!$ arises upon
counting all possible ways of ordering the $n$ loops.
Diagrams which generalize the leading order contribution to the
variance for $n=3$ are shown in Fig.~\ref{fig:5}.


\ifpreprintsty \else
			\end{multicols}\vspace*{-3.5ex}{\tiny
                   \noindent\begin{tabular}[t]{c|}
                   \parbox{0.493\hsize}{~} \\ \hline \end{tabular}}
\vspace*{3.5ex}
\begin{figure}
\vspace{0.3cm}
\hspace*{5mm}
\epsfxsize=0.9\hsize
\epsffile{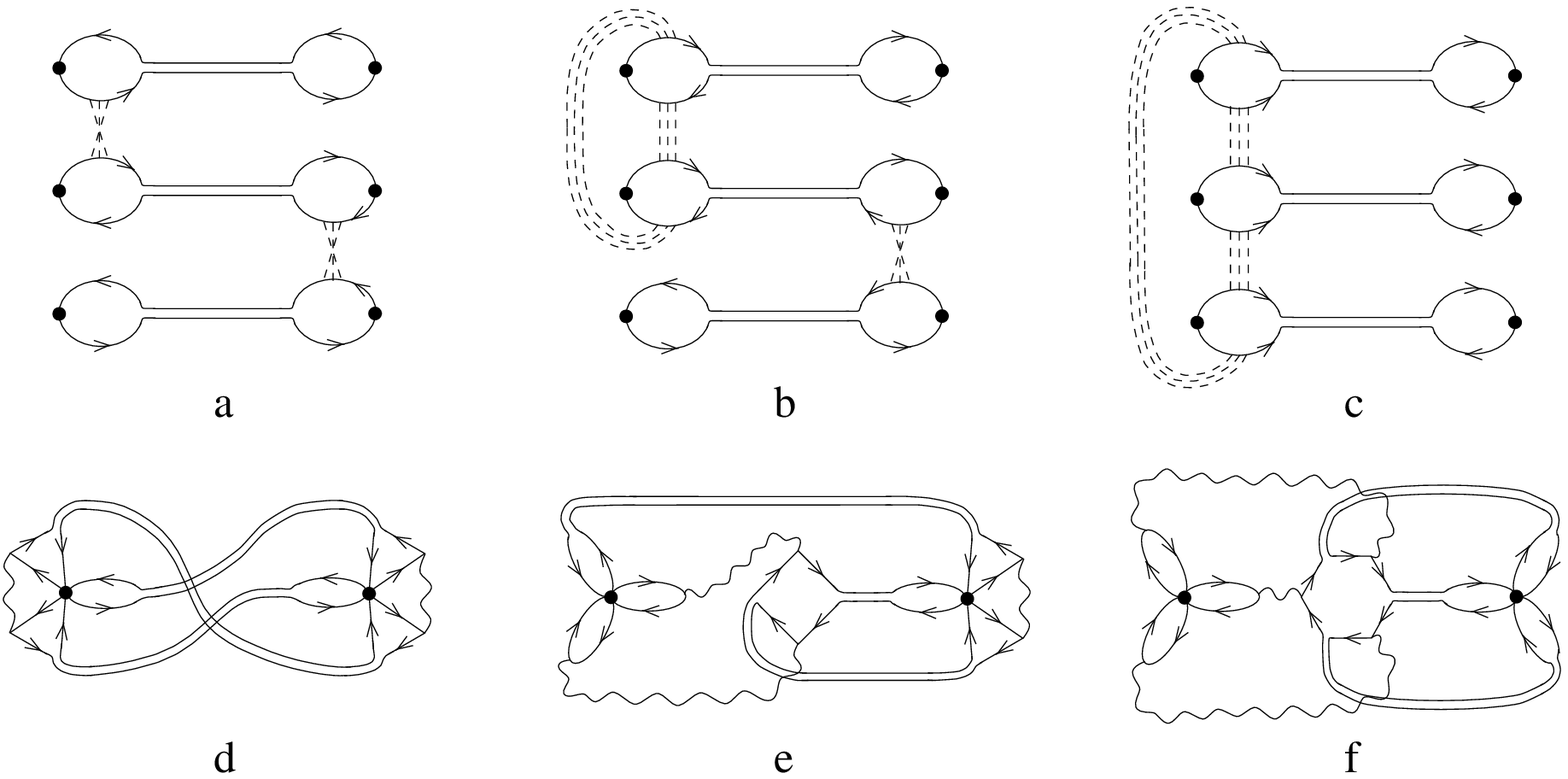}
\vspace{0.3cm}
 
\refstepcounter{figure}
\label{fig:5}
{\small \setlength{\baselineskip}{10pt} FIG.\ \ref{fig:5}.
Some of the diagrams making a leading contribution
to the third moment of conductance. Diagrams in the second row are 
equivalent to their counterparts in the first row. 
}
\end{figure}
\begin{multicols}{2}\noindent \fi

In general, such diagrams are obtained by 
taking $n$ conductance loops, each
loop having an intra-sample diffuson (as for the mean, Fig.~2), and connecting
these loops with the least possible number of inter-sample diffusons.
Actually, the number of inter-sample diffusons may vary, as every 
inner polygon (like the inner  square in diagram \ref{fig:4}{\it d}
for the variance) cancels one of them (see Appendix).
It is the number of loops consisting
of inter-sample diffusons which should be minimal.

For the third cumulant (Fig.~5.),
{each diagram} has two such loops.
The lowest order diagram,  two  three-petal daisies 
connected by three inter-sample diffusons, also has two diffuson loops. 
However, there is no cancellation there and this diagram is proportional
to $\zeta_\phi(R)^3$. Each diagram in Fig.~5 is proportional to
$\zeta_\phi^2(\ell) \zeta_{\text{esc}}^3$. As we are interested in the case
of strong dephasing ($\zeta_\phi\ll1$) and weak leakage $\zeta_{\text{esc}}
\sim1$, this contribution is dominant. Note that Fig.~5 does not
include all the diagrams of this order, and a numerical factor 
attached to the leading contribution to 
$\left<\!\left<G^n\right>\!\right>$ is not easy to calculate for
$n=3$ and hardly possible for arbitrary $n$. However, 
this is not necessary for what follows. It is sufficient to know
that the leading contribution to the $n$th cumulant is proportional
to $\zeta_{\phi}^{n-1}(\ell )\,\zeta_{\text{esc}}^n$. 
Thus, keeping both the lowest
order and the leading contribution, we find
\begin{equation}
\frac{\left<\!\left<G^n\right>\!\right>}
{\left<G\right>^n} 
 = \frac{(n-1)!}{\beta}\left\{\left[
{\zeta_{\phi}(R)\over{\zeta_{\text{esc}}(R)}}\right]^n
+ \, c_n \,\,\left[ \zeta_{\phi}(\ell )\right]^{n-1}\right\},
\label{pointn}
\end{equation}
where $n$-dependence of $c_n$ is much slower 
than factorial (although it might include a trivial factor
like $2^n$ or so). For weak particle escape and strong dephasing,
Eq.\ (\ref{ineq}), the second  term in Eq.\ (\ref{pointn}) dominates. 
 Comparison of this dominant contribution 
to an appropriate power of the variance gives
$$\frac{\left<\!\left<G^{n}\right>\!\right>^2}
{(\text{var}\, G)^n} \sim 
\left[ \zeta_{\phi}(\ell )\right]^{{n-2}}\ll1\,,$$
so that the distribution
is mainly Gaussian with non-universal variance 
$\text{var}\, G$ given  by Eqs.\ (\ref{var}) and
(\ref{vg2}). Note that this is similar to the behavior of the higher
order cumulants \cite{AKL:86} in the case of UCF in a sample
with broad contacts, where such a ratio is proportional to 
$g_0^{-2(n-2)}$. Very high cumulants, with $n\agt g_0^{-1}$, grow much faster
which leads to lognormal tails of the distribution\cite{M+L:96} 
similar to those \cite{AKL:86} for samples with broad contacts.
At weak disorder the lognormal tails are practically irrelevant as they
start  at $\delta G \sim \left<G\right>
\sqrt{g_0}\gg \left<G\right> $	where $\delta G =G-\left<G\right>$, 
while typical deviations $\delta G \sim \sqrt{\text{var}\,G}\ll \left<G\right>
$.

In the presence of some particle leakage from the dot,
$\zeta_{\text{esc}}(R)\agt 1$,  the exponential tails are more important, 
these tails being governed by  first term in Eq.\ (\ref{pointn}) which
dominates for larger $n$. 
It turns out that the
exponential tails dominate the Gaussian ones for
\begin{equation}
\left(\delta G \right)^2 \gg  \text{var}\,  G\,
\frac{ \zeta_{\text{esc}}^2} {\zeta_{\phi}}\,.
 \label{expt}
\end{equation}  
\ifpreprintsty \else
\begin{figure}
\hspace{0.05\hsize}
\epsfxsize=0.9\hsize
\epsffile{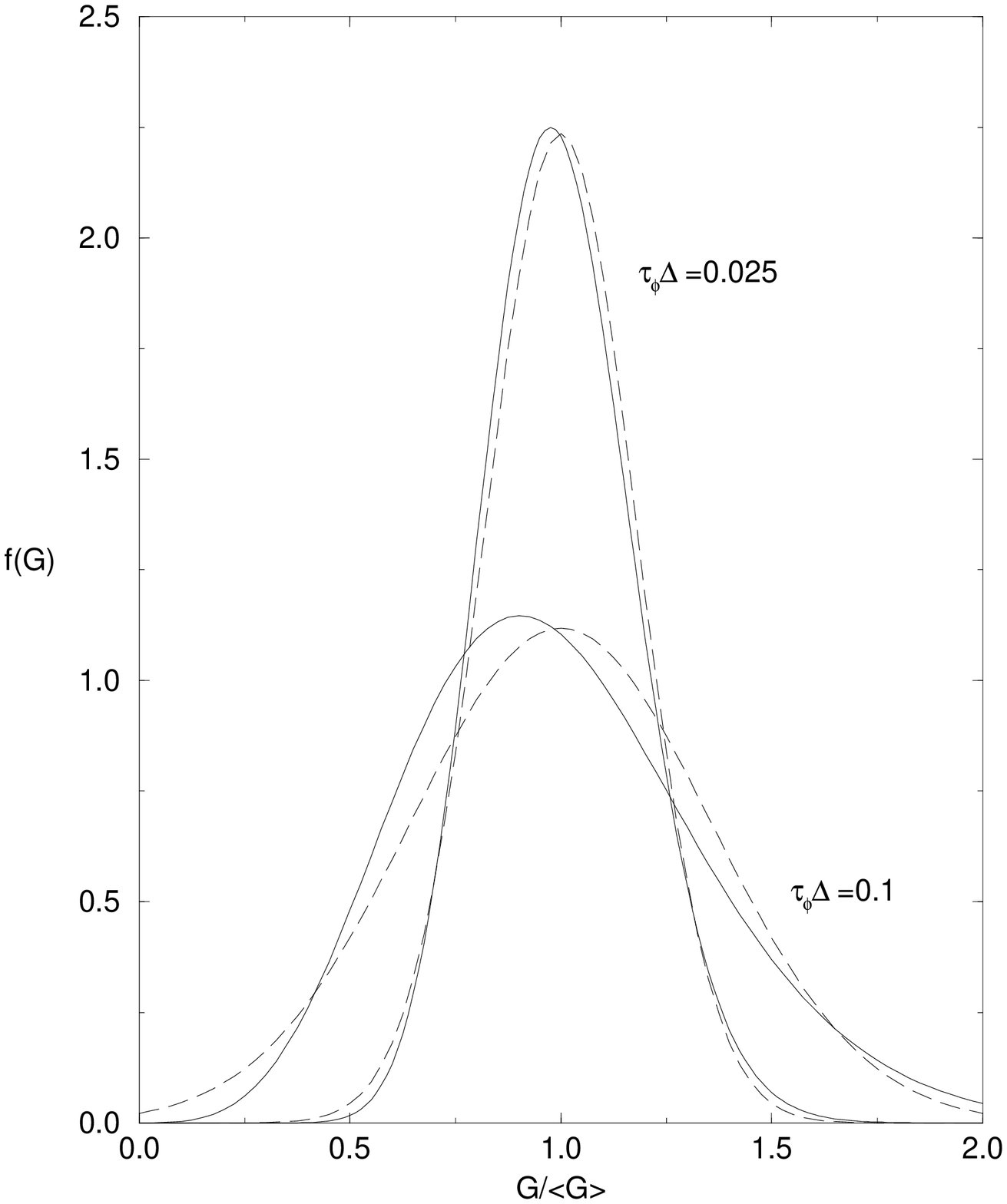}
\vspace{0.3cm}
 
\refstepcounter{figure}
\label{fig:6}
{\small \setlength{\baselineskip}{10pt} FIG.\ \ref{fig:6}.
Solid curves show the conductance distribution for $\beta =1$ 
with dephasing rates $\Delta\tau_{\phi} = 0.1$ and $0.025$, 
calculated using the second term in Eq.\ (\ref{pointn}) 
with an arbitrarily chosen coefficient $c_n = 2^{n-1} n$. 
For comparison the dashed curves are Gaussian distributions 
with the same mean and variance.
}
\end{figure}\noindent\fi 
This estimation results from
 a comparison of the exponential distribution obtained using the
first term in Eq.~(\ref{pointn}) with the Gaussian distribution with 
the variance (\ref{vg2}) obtained by neglecting the higher cumulants.

For  weak particle leakage, the exponential tails start at deviations
which are much stronger than typical ones. With increasing leakage,
these tails become important at the same values of $\zeta_{\text{esc}}$
at which the variance  is mainly contributed by escape processes,
Eq.~(\ref{ratio}). 
Inequalities (\ref{ratio}) and  (\ref{expt}) 
show that the crossover to the regime dominated
by particle escape happens not at $\tau_{\text{esc}}\sim
\max(\tau_\phi, \tau _{\text{erg}})$ but at considerably weaker
particle escape: either at
$\tau_{\text{esc}}\sim \sqrt{\tau_\phi/\Delta}\gg\tau_\phi$ in the
ergodic regime ($\tau_\phi\gg\tau _{\text{erg}}$), or at
$\tau_{\text{esc}}\sim\sqrt{\tau_{\text{erg}}/\Delta}\gg\tau_{\text{erg}}$
in the diffusive regime ($\tau_\phi\ll\tau _{\text{erg}}$).
Note that in Eqs.~(\ref{ratio}) and  (\ref{expt}) we have disregarded
{the} difference between $\zeta_\phi(R)$ and $\zeta_\phi(\ell)$ which is
exact in the zero-mode regime and logarithmically accurate in the 
diffusive regime at $d=2$. In the diffusive regime at $d=3$, 
this difference is significant as may be seen from Eq.\ (\ref{zr}). Taking
this into account will introduce an extra small factor of $\ell/R$ 
into the r.h.s.\ of the estimations (\ref{ratio}) and  (\ref{expt})
thus suppressing a relative importance of escape processes. 

In the case of main interest, when particle leakage may be neglected,
the distribution still deviates from Gaussian since the cumulants, given
by the second  term in Eq.\ (\ref{pointn}), do not  vanish although they
are small. The distribution can be restored from these cumulants in 
a standard way (see, e.g., Ref.\ \onlinecite{AKL:86}). Exact values of 
the coefficient $c_n$ do not matter as the most relevant {contribution}
to the deviation from a Gaussian shape is given by the third cumulant,
similarly to the case of the conductance fluctuations in samples
with broad leads.\cite{vRLAN} 	The resulting distributions are
shown for two values of $\tau_\phi/\tau_{\text{erg}}$ in Fig.~6. 
It is seen that deviations from the Gaussian shape are not very strong.

In conclusion,
we have found that the conductance distribution for disordered quantum
dots with single-channel leads becomes mainly Gaussian for
strong dephasing, $\tau_\phi\ll h/\Delta$, 
in broad agreement with the voltage probe model 
of Ref.~\onlinecite{B+B:97}.
Let us stress again that 
since  we have assumed non-zero $\gamma_{\text{leak}}$ in this
paper, there is no exact region of overlap between our results and those of the
voltage probe model.
Therefore, a direct comparison of the results is not possible. 
 However there remain some noticeable differences, 
first of all in the dependence of the variance of the
Gaussian distribution on $\tau_\phi$.
If we stretch our results to the limit of applicability putting
$\zeta_{\text{esc}}=1$ and making the contacts transparent ($\alpha_1=\alpha_2
=1$), we will have var$G \sim \tau_\phi$ while the result of
Ref.~\onlinecite{B+B:97} for transparent contacts is var$G\sim  \tau_\phi^2$. 
Note that although in our approach resummation in all powers of 
$\zeta_{\text{esc}}$
might be necessary for $\zeta_{\text{esc}}\alt 1$, it could not change the 
dependence of the result on $\tau_\phi$. Similarly, in contrast to another
result of Ref.~\onlinecite{B+B:97}, the dephasing rate $\tau_\phi^{-1}$
could not appear in the mean conductance in the unitary case, $\beta =2$. 
Indeed, in this case Cooperons vanish, and particle conservation 
 strictly prohibits the appearance
of $\tau_\phi^{-1}$ in any power of expansion in diffusons. 
However, this disagreement is minor, and is clearly due to differences
 between the models. Let us stress that in contrast to the imaginary potential
 model, our present microscopic considerations confirm the prediction 
 of the voltage probe model on the Gaussian character of the 
 conductance distribution with a non-universal variance depending
 on the dephasing rate.

 The most distinctive feature
of the perturbative approach used here was taking into account two different
sources of level broadening, one due to the presence of contacts
entering the classical part of the mean
conductance, and the other due to dephasing, appearing only in
weak localization corrections and fluctuations.
 It remains an open question as to whether a similar introduction of two
different sources of level broadening could be used within a non-perturbative
approach such as the supersymmetric nonlinear $\sigma$ model, thus enabling a
calculation for $\gamma_{\text{leak}} =0$.

\acknowledgments
We are grateful to C.~W.~J.~Beenakker, {M. Leadbeater}, and R.~A.~Smith for 
useful 
discussions.
Work in Birmingham has been supported by 
EPSRC grants GR/J35238 and  GR/K95505. 

\appendix
 \section{Calculation of the leading contribution to the variance}
 
 The  calculation of the diagrams in Fig.~\ref{fig:4} is straightforward
in the representation ({\it b}) and ({\it d}). The contribution of 
diagram ({\it b}) is given by
\begin{equation}
8A^2 |\chi_3|^2 \chi_2^2 {\cal D} 
(\bbox{r}_1, \bbox{r}_1;\,i\gamma_\phi) {\cal D}^2
(\bbox{r}_1, \bbox{r}_2;\,i\gamma_{\text{esc}})
\label{4b}
\,.\end{equation}
Here the overall factor of $8$ takes into account that diagram (4{\it b}) 
originates from $\left<T^LT^R\right>$, and there are four ways to connect
$T^R$ and $T^L$ with an inter-sample diffuson which is clear from the 
equivalent representation of this diagram, Fig.~\ref{fig:4}{\it a}.
The constant
 $\chi_3$ is contributed by a triangle consisting of three
Green's functions. This is a constant since,
similarly to the calculation of the mean conductance,
Eq.~(\ref{meang}), the triangle  may be approximated
 at the diffusive scale, $| {\bf  r} -
{\bf  r^{\prime}}| \agt \ell$, by
$\chi_3( {\bf  r} , {\bf  r^{\prime}}  , {\bf  r^{\prime\prime}})=
\chi_3\,\delta({\bf  r} - {\bf  r^{\prime}})
\,\delta({\bf  r} - {\bf  r^{\prime\prime}})$. The presence of the
$\delta$  functions ensures that the contribution of this diagram, 
Eq.\ (\ref{4b}), reduces to a mere product of the local
contributions, $\chi_2$ 
and $\chi_3$, and the diffusons,
 where  the constant $\chi_3 $ is given by
\begin{equation}
\chi_3=
\int \left<{\cal G}^{+}( {\bf p})\right>^2
\left<{\cal G}^{-}( {\bf p})\right>{\rm d}^d p
=-2\pi i\nu_0\tau^2\,.
\label{chi3}
\end{equation}
At the same scale, the spatial arguments in ${\cal D}
(\bbox{r}_1, \bbox{r}_1;\,i\gamma_\phi)$ coincide within an accuracy
of order  $\ell$. Therefore,
 $2\tau^2 {\cal D}
(\bbox{r}_1, \bbox{r}_1;\,i\gamma_\phi)=
\zeta_{\phi} (R\!=\!\ell)\equiv \zeta_{\phi}(\ell )$
 by the definition of Eq.~(\ref{dr}) with
$\zeta_{\phi} (R)$ given by  Eq.~(\ref{zr}).
Substituting all the constants into Eq.~(\ref{4b}), we obtain the contribution
of diagram ({\it b}) as follows
\begin{equation}
\left[{e^2\alpha_{1}
\alpha_{2} \over{h}}
\zeta_{\text{esc}}(R)\right]^2 \!\! \zeta_{\phi}(\ell )
= {\zeta_{\phi}(\ell )}\,\left<G\right>^2
\,.
\label{diagb}
\end{equation}

The contribution of diagram  (4{\it d}) is given by
\begin{equation}
4A^2 \chi_2^4 {\cal D}^2 (i\gamma_{\text{esc}})
 \int\!\hat\chi_4(\bbox{r})\,{\cal D}^2 (\bbox{r_1},\bbox{r};i\gamma_\phi)
{\rm d}^dr\,.
\label{4d}
\end{equation}
Here $\chi_4$ is contributed by the ``dressed'' inner
square which is the sum of one ``bare'' square and two squares containing
one single-impurity line each, this line connecting opposite sides of the
square.\cite{Gor:79}
The factor of 4 arises because diagram (4{\it d})
originates from $\bigl<\bigl(T^L\bigr){}^2\bigr>+
\bigl<\bigl(T^R\bigr){}^2\bigr>$ and there are only two ways to connect
the conductance loops with two inter-sample diffusons which is clear
from the equivalent representation, Fig.~\ref{fig:4}{\it c}.

The calculation is more  transparent in the
ergodic zero-mode regime, $\tau_{\text{erg}}\ll \tau_\phi \ll \hbar
/\Delta$, when not only the intrinsic diffuson, Eq.\ (\ref{zr}),
but also the inter-sample diffuson, Eq.\ (\ref{esc}), and the
dressed square, $\chi_4$, reduce to a constant. 
Since the  contribution of a single-impurity line 
is given by the correlator (\ref{RP}), one finds $\chi_4$ as follows
$$
\chi_4=
\int \!\!\left<{\cal G}^{+}\right>^2
\left<{\cal G}^{-}\right>^2{\rm d}^d p
+\frac1{2\pi \nu_0\tau_{\text{el}}}
\left(\int \!\!\left<{\cal G}^{+}\right>^2
\left<{\cal G}^{-}\right>{\rm d}^d p\right)^2
$$
The first term is calculated in a similar manner to that in Eq.~(\ref{chi2})
which gives $4\pi\nu_0\tau^3$. The integral in the second term
coincides with that in Eq.~(\ref{chi3}) so that one obtains 
\begin{equation}
\chi_4=4\pi\nu_0\tau^4\left(\frac1\tau-\frac1{\tau_{\text{el}}}\right)
=\frac{4\pi\nu_0\tau^4}{\tau_\phi}\,.
\label{chi4}
\end{equation}
The difference between the one-particle relaxation time, $\tau$, and 
the mean elastic scattering time, $\tau_{\text{el}}$, is crucial 
in this calculation; the same difference is a formal cause of the 
saturation of the inter-sample diffuson, Eq.\ (\ref{dr}),
for $q=0$ at $\gamma=1/\tau_\phi$. 

On substituting the zero-mode values of $\chi_4$, Eq.\ (\ref{chi4}), 
and ${\cal D}(i\gamma_\phi$), one reduces the integral in Eq.\ (\ref{4d})
to 
\begin{equation}
 \int\!\hat\chi_4(\bbox{r})\,{\cal D}^2 (\bbox{r_1},\bbox{r};i\gamma_\phi)
{\rm d}^dr=\frac{4\pi\nu_0\tau^4}{\tau_\phi}\,
\frac{L^d\zeta_\phi^2}{(2\tau^2){}^2}=\zeta_\phi\,,
\label{int1}
\end{equation}
so that the dressed square, $\chi_4$, exactly cancels one of the inter-sample 
diffusons. Then, on substituting Eq.\ (\ref{int1}) and all the constants
into  Eq.\ (\ref{4d}), one finds that in the zero-mode case (when
$\zeta_\phi(\ell)\equiv\zeta_\phi=\Delta\tau_\phi/\pi$), this contribution
equals  that of diagram (4{\it b}),  Eq.\ (\ref{diagb}). In  Eq.\ (\ref{chi4})
we have neglected the $1/\tau_{\text{esc}}$ contribution to the one-particle
relaxation time, $1/\tau$. Taken into account, this would lead to an
additional contribution of diagram ({\it d}) proportional to
$\zeta_\phi^2\zeta_{\text{esc}}$ rather than $\zeta_\phi\zeta_{\text{esc}}^2$
as in Eq.\ (\ref{diagb}). In this contribution, the inner square cancels
one of the intrinsic diffusons. Overall, such a contribution of diagram ({\it 
d})
would be exactly equal to the contribution of the diagram similar
to that in Fig.~(\ref{fig:4}{\it b})
but made from one intrinsic and two inter-sample diffusons. We neglect
these contributions which are small even compared to those of the diagrams
in  Fig.~(\ref{fig:3}).

We do not present here the calculation of $\hat \chi_4(\bbox{r})$ in a
general case, without the restriction to the zero-mode regime.
The calculation is standard\cite{Gor:79} and results in 
$$
\hat \chi_4(\bbox{r})=
{2\pi\nu_0\tau^4}\Bigl(-D\nabla^2+\gamma_\phi\Bigr)\,,
$$
so that $\hat \chi_4(\bbox{r})$ is proportional to the diffusion operator.
Then using the diffusion equation one finds that
$$
 \int\!\hat\chi_4(\bbox{r})\,{\cal D}^2 (\bbox{r_1},\bbox{r};i\gamma_\phi)
{\rm d}^dr=2\tau^2 {\cal D} (\bbox{r_1},\bbox{r}_1;i\gamma_\phi)
=\zeta_\phi(\ell)\,,
$$
which generalizes Eq.\ (\ref{int1}) and shows that the contribution of 
diagram (\ref{fig:4}{\it d}) in this case is also exactly equal to that 
of diagram (\ref{fig:4}{\it b}),  Eq.\ (\ref{diagb}).

Finally, the overall coefficient for the contribution of all the leading
diagrams equals $4/\beta$, since in the orthogonal case ($\beta=1$) an
equal contribution is made by two equivalent Cooperon diagrams. This leads to
the result of Eq.\ (\ref{vg2}).

\ifpreprintsty
\begin{figure}
\vspace*{1.3cm}
\hspace{0.02\hsize}
\epsfxsize=0.9\hsize
\epsffile{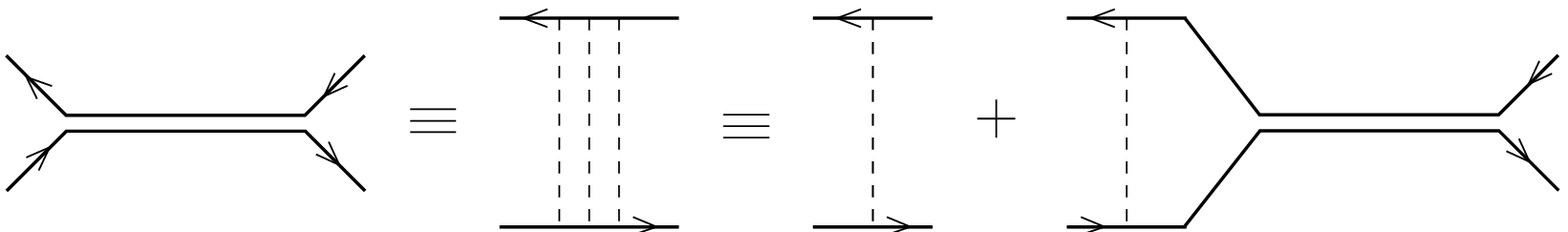}
\vspace{0.3cm}
 
\refstepcounter{figure}
\label{fig:1}
{\small \setlength{\baselineskip}{10pt} FIG.\ \ref{fig:1}.
A diffuson ladder. A Cooperon ladder is obtained by inverting
the direction of one of {the} arrows. }
\end{figure}
 
\vfil
\begin{figure}
\vspace{0.3cm}
\hspace{0.12\hsize}
\epsfxsize=0.6\hsize
\epsffile{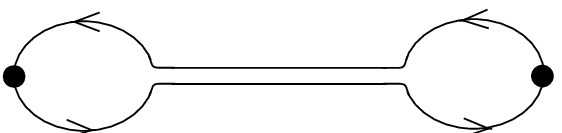}
 
\vspace{0.3cm}
\refstepcounter{figure}
\label{fig:2}
{\small \setlength{\baselineskip}{10pt} FIG.\ \ref{fig:2}.
The diagram for the mean conductance.}
\end{figure}
\vfil

\newpage
\begin{figure}
\vspace{0.3cm}
\epsfxsize=0.95\hsize
\epsffile{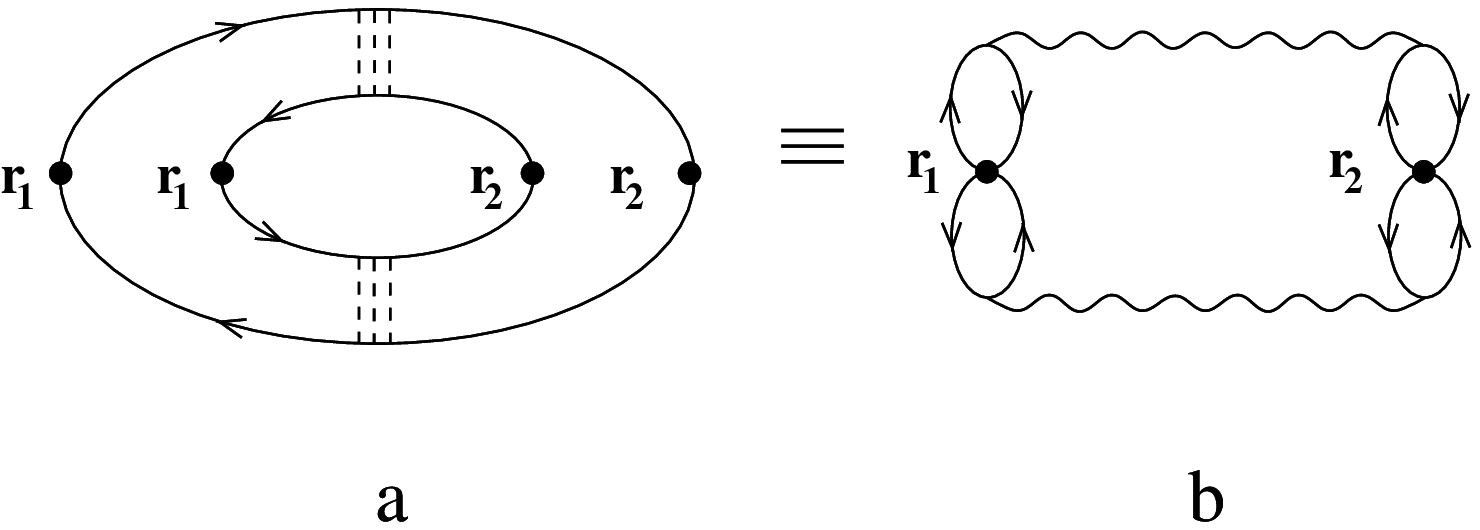}
\vspace{0.3cm}
 
\refstepcounter{figure}
\label{fig:3}
{\small \setlength{\baselineskip}{10pt} FIG.\ \ref{fig:3}.
The lowest-order contributions to the variance. Wavy lines represent
{diffuson} ladders which correspond to the {inter-sample} diffusons. The
relation between a wavy line and a ladder is the same as for intrinsic
diffusons, Fig.\ \ref{fig:1}.
 }
\end{figure}
\vfill\break

\begin{figure}
\vspace{0.3cm}
\epsfxsize=0.95\hsize
\epsffile{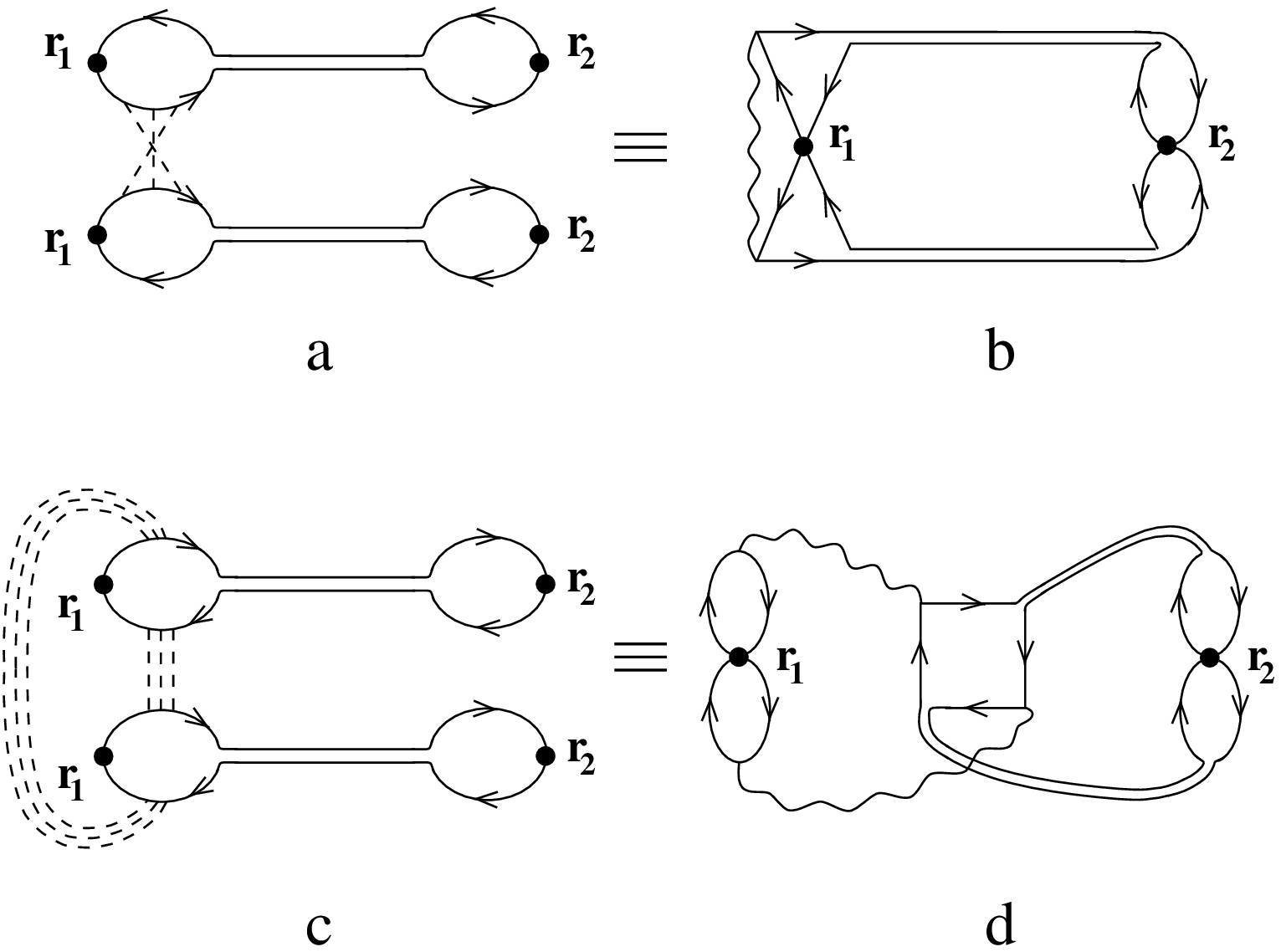}
\vspace{0.3cm}
 
\refstepcounter{figure}
\label{fig:4}
{\small \setlength{\baselineskip}{10pt} FIG.\ \ref{fig:4}.
The leading diagrammatic contribution to the variance. 
In the diagrams {\em b} and {\em d}, the inter-sample diffusons
are represented by wavy lines and the intrinsic diffusons by double
lines. 
}
\end{figure}
\newpage
\begin{figure}
\vspace{0.3cm}
\hspace{0.02\hsize}
\epsfxsize=0.9\hsize
\epsffile{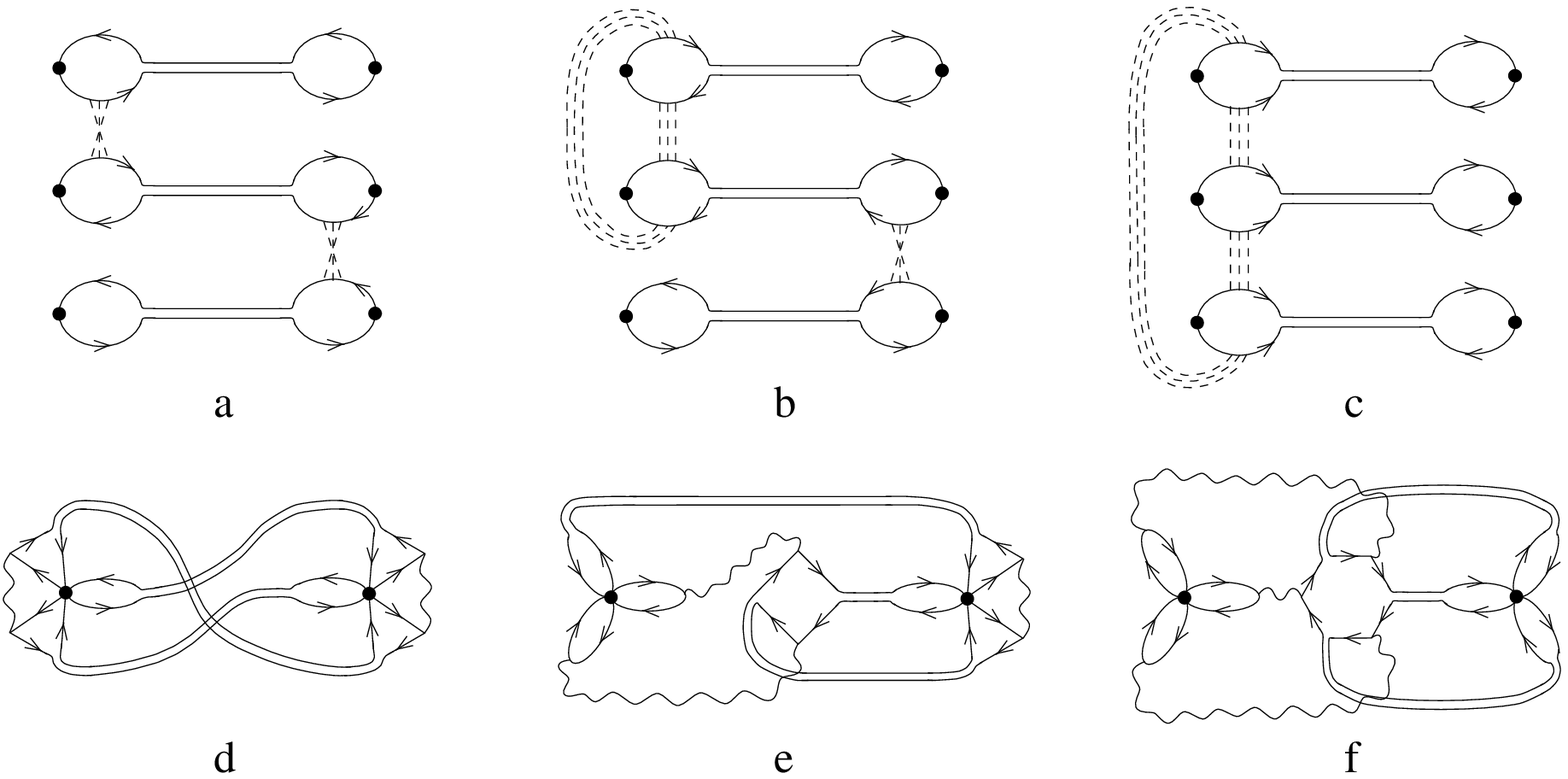}
\vspace{0.3cm}
 
\refstepcounter{figure}
\label{fig:5}
{\small \setlength{\baselineskip}{10pt} FIG.\ \ref{fig:5}.
Some of the diagrams making a leading contribution
to the third moment of conductance. Diagrams in the second row are
equivalent to their counterparts in the first row. 
}
\end{figure}

\begin{figure}
\vspace{0.3cm}
\hspace{0.05\hsize}
\epsfxsize=0.9\hsize
\epsffile{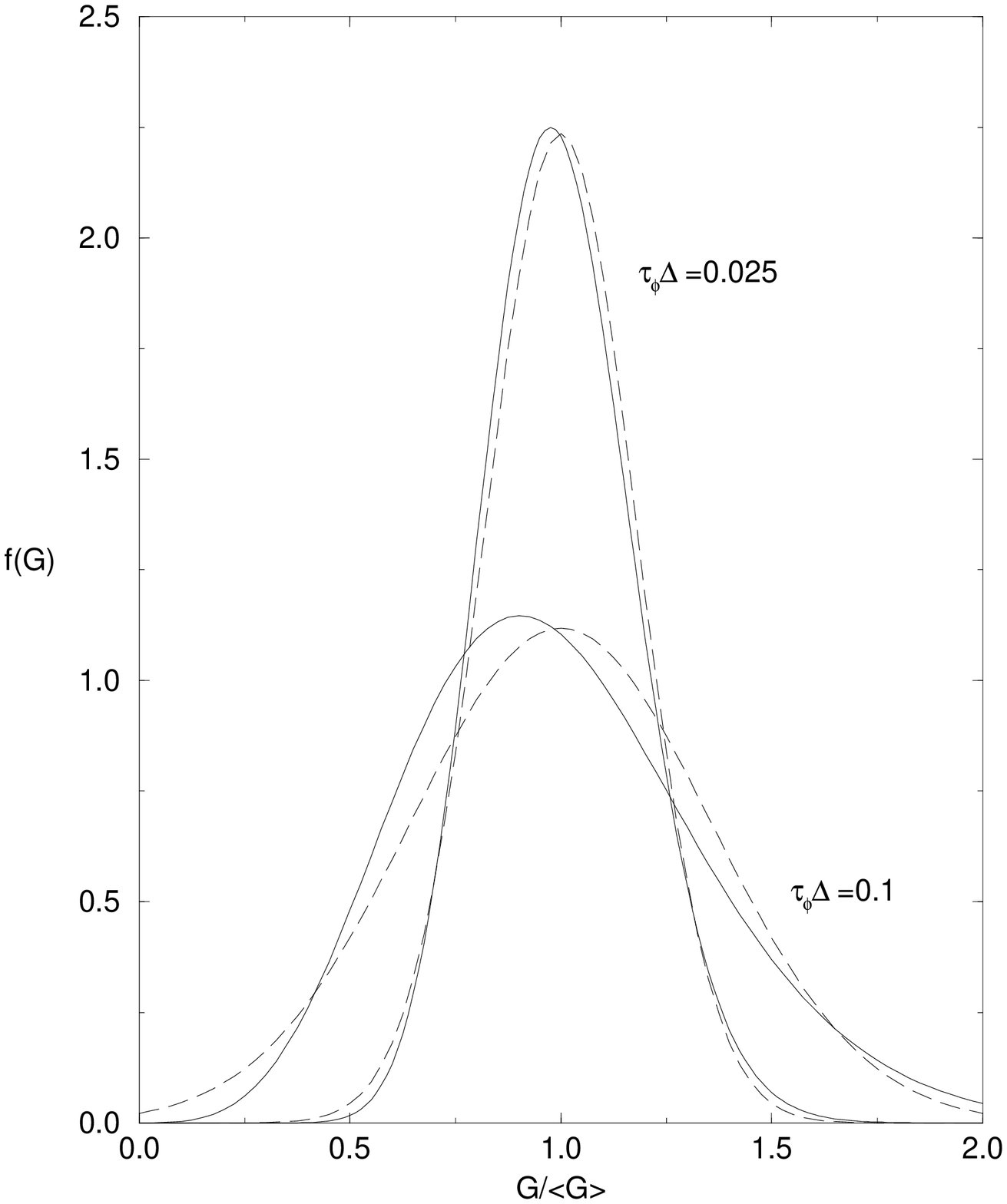}
\vspace{0.3cm}
 
\refstepcounter{figure}
\label{fig:6}
{\small \setlength{\baselineskip}{10pt} FIG.\ \ref{fig:6}.
Solid curves show the conductance distribution for $\beta =1$ 
with dephasing rates $\Delta\tau_{\phi} = 0.1$ and $0.025$, 
calculated using the second term in Eq.\ (\ref{pointn}) 
with an arbitrarily chosen coefficient $c_n = 2^{n-1} n$. 
For comparison the dashed curves are Gaussian distributions 
with the same mean and variance.
}
\end{figure}
                   \else \end{multicols} \fi

\begin{references}
\bibitem{tau-el}
Average Green's functions $\left<{\cal G}^{\pm}( \xi )\right>=(\xi\pm
  i/2\tau)^{-1}$ include $\tau^{-1}=\tau_{\text{el}}^{-1}+\tau_\phi^{-1}\approx
  \tau_{\text{el}}^{-1}$. In calculating petals or any boxes containing filled
  circles (i.e.\ representing electron motion at the ballistic scale near the
  leads), one can safely substitute $\tau_{\text{el}}$ for $\tau$. However, the
  difference between $\tau_{\text{el}}$ and $\tau$ is crucial in calculating
  `dressed' inner boxes without external vertices like the square in diagram 
4$d$,
see Appendix.
\end{references}
\end{document}